\documentstyle[onecolumn]{mn}
\newcommand{\mincir}{\raise -2.truept\hbox{\rlap{\hbox{$\sim$}}\raise5.truept
\hbox{$<$}\ }}
\newcommand{\magcir}{\raise -2.truept\hbox{\rlap{\hbox{$\sim$}}\raise5.truept
\hbox{$>$}\ }}
\newcommand{\minmag}{\raise-2.truept\hbox{\rlap{\hbox{$<$}}\raise 6.truept\hbox
{$>$}\ }}
\newcommand{\be}{\begin{equation}}
\newcommand{\ee}{\end{equation}}
\newcommand{\ba}{\begin{eqnarray}}
\newcommand{\ea}{\end{eqnarray}}
\newcommand{\brr}{\begin{array}}
\newcommand{\err}{\end{array}}
\newcommand{\bc}{\begin{center}}
\newcommand{\ec}{\end{center}}

\title{Redshift evolution of clustering}
\author[Matarrese et al.]
{Sabino Matarrese,$^{1}$ Peter Coles,$^{2}$ Francesco Lucchin$^{3}$ and
Lauro Moscardini$^{3}$\\
$^1$Dipartimento di Fisica G. Galilei, 
Universit\`{a} di Padova, via Marzolo 8, I--35131 Padova, Italy\\
$^2$Astronomy Unit, School of Mathematical Sciences,
Queen Mary \& Westfield College, Mile End Road, London E1 4NS\\
$^3$Dipartimento di Astronomia, Universit\`a di Padova,
vicolo dell'Osservatorio 5, I--35122 Padova, Italy}
\begin{document}

\maketitle

\begin{abstract}
We discuss how the redshift dependence of the observed two--point correlation 
function of various classes of objects can be related to theoretical 
predictions. 
This relation involves first a calculation of the redshift evolution
of the underlying matter correlations. The next step is to relate
fluctuations in mass to those of any particular class of cosmic
objects; in general terms, this means a model for the bias and how
it evolves with cosmic epoch. Only after these two effects have been
quantified can one perform  an appropriate convolution of the 
non--linearly evolved  two--point correlation 
function of the objects with their redshift 
distribution to obtain the `observed' correlation function for 
a given sample. This convolution in itself
tends to  mask the effect of evolution by mixing amplitudes
at different redshifts. We develop a formalism which incorporates
these requirements and, in particular, a set of plausible models
for the evolution of the bias factor. We apply this formalism to the 
spatial, angular and projected correlation functions from 
different samples of high--redshift objects, assuming 
a simple phenomenological model for the initial power--spectrum
and an Einstein-de Sitter cosmological model.
We find that our model is roughly consistent with data on the
evolution of QSO and galaxy clustering, but only if the effective
degree of biasing is small. We discuss the differences between our analysis
and other theoretical studies of clustering evolution and argue
that the dominant barrier to making definitive predictions is
uncertainty about the appropriate form of the bias and its
evolution with cosmic epoch.
\end{abstract}

\begin{keywords}
cosmology: theory -- cosmology: observations -- large--scale structure
of Universe -- galaxies: formation -- galaxies: evolution -- galaxies: haloes
\end{keywords}

\section{Introduction}

The analysis of the clustering properties of various classes of objects at 
high redshift is rapidly becoming one of the key tests of models for the
formation and present properties of the large--scale structure of the 
universe. The importance of such studies is that they offer the
prospect of highlighting differences between viable theoretical models which,
by construction, display clustering behaviour which is
very similar at the present epoch (e.g. Coles 1996). Improvements
in observational technology also offer the prospect of providing
strong constraints on the nature of any bias that exists between the
clustering of objects of a particular type and that of the underlying matter 
distribution. The possible existence of a significant bias in the
galaxy distribution is a presently a significant obstacle, for example, to the
determination of the cosmological density parameter $\Omega_0$
from large--scale clustering data (e.g. Coles \& Ellis 1996).
This second point is particularly relevant because of the
wide range of different objects which are now accessible to 
systematic redshift survey programmes - not only the `traditional' bright, 
optically selected galaxies upon which
surveys were based in the 1980s, but now also faint optical
galaxies, galaxies selected in the infrared, QSOs, absorption--line systems,
radio-galaxies and clusters of galaxies
selected either optically or through their X-ray emission. Each of these
classes of objects might be related to the matter distribution in a
different way, a possibility that, on the one hand, poses great
difficulties of interpretation but, on the other, admits the possibility
of realising many independent tests of theories for the formation of
structure. 

In spite of the many important recent observational
developments,  theoretical models used to 
interpret the data have so far generally been naive and poorly motivated 
in terms of galaxy formation scenarios. This theoretical inadequacy is 
clearly shown by the wide use, primarily  in the literature  
relating to observational data, of the model 
\be
\xi(r,z) = \xi(r/(1+z),0) (1 + z)^{-(3+\epsilon)} 
\label{eq:theor}
\ee
for the redshift evolution of the two--point correlation function $\xi(r,z)$ 
at the comoving separation $r$, where $\epsilon$ is an arbitrary fitting 
parameter. If the space dependence of the two--point function can be 
fitted by a power law, the above relation simplifies to 
\be
\xi(r,z) = (r/r_c)^{-\gamma} (1+z)^{-(3-\gamma+\epsilon)} \;,
\label{eq:powlaw}
\ee
where $r_c$ is a constant measuring the unity crossing of $\xi$ at $z=0$. 

Recent observational studies have served to highlight the importance
of understanding the validity (or otherwise) of the simple models
(1) \& (2). For example, using equation (\ref{eq:theor}) 
as a fitting formula for the two--point function of faint 
galaxies, various authors (e.g. Efstathiou et al. 
1991; Neuschaefer, Windhorst \& Dressler 1991; Couch, Jurcevich \& Boyle 1993;
Roche et al. 1993; Bernstein et al. 1994; Cole et al. 1994; 
Brainerd, Smail \& Mould 1995; Le F\`evre et al. 1996; Roche et al. 1996;
Shepherd et al. 1996) have found 
observational evidence for an excess evolution with respect to the stable 
clustering prediction ($\epsilon=0$), 
i.e. for positive values of $\epsilon$ (the so--called 
{\em collapsing models}).  In particular, Shepherd et al. (1996) and 
Le F\`evre et al. (1996), from the 
analysis of galaxy clustering at moderate redshifts, obtained $\epsilon 
\sim 1 \pm 1$. But is this a real (physical) discrepancy with realistic
models, or would more sophisticated modelling of the evolution of
$\xi(r,z)$ yield a different interpretation? We incline to the second view,
for reasons we explore below, and suggest that the quantity and quality of the high-redshift data now
available requires a significant improvement in the mathematical models
deployed for their interpretation.

In some ways this situation is reminiscent of the way some 
pre--COBE analyses of the CMB angular 
two--point function were carried out: in most experimental papers the data were fitted in 
terms of the so--called `Gaussian correlation function', which had no 
justification in terms of theoretical models for the temperature anisotropy 
pattern. Likewise these simple power law models for the correlation
function have relatively little theoretical motivation and a more
realistic approach is consequently required.

Assuming that clustering grows by gravitational instability, the above formula 
is only justified in two cases: for $\epsilon=0$ it reproduces the prediction 
of the so--called {\em stable clustering} model, while, for 
$\epsilon=n+2=\gamma-1$, it results from the application of linear theory in 
an Einstein--de Sitter universe to purely scale--free power--spectra, 
$P_{\rm lin}(k,0) \propto k^n$. The case where $\epsilon=\gamma-3$ 
corresponds to a clustering pattern that simply expands with the background
cosmology as if the galaxies were just painted on a homogeneous
background. Concerning the first case, one should remember that 
the idea underlying the stable clustering ansatz is that, on 
sufficiently small scales, gravity acts to stabilize the number of 
neighbours of an object in a proper volume, after this has turned around from 
the universal expansion. Although the physical grounds of 
this model appear to be reasonably sound (e.g. Jain \& Bertschinger 1996 and 
references therein; see, however, Padmanabhan et al. 1996), numerical 
simulations indicate that this type of dynamics is only 
relevant on scales where the mass autocorrelation function is at least 
as large as $\sim 100$ (e.g. Efstathiou et al. 1988), i.e. at most a small 
fraction of the typical scales probed by catalogues. The second model is also 
inadequate to treat the behaviour of correlation functions in any realistic 
galaxy formation scenario.

Melott (1992) has analysed the growth of clustering in numerical simulations 
for an ensemble of scale--free models. He finds that the lower is the 
value of the spectral index $n$, the larger is the value of the parameter 
$\alpha\equiv\ 3 - \gamma + \epsilon$ and that positive values of $\epsilon$ 
are easily allowed for in all models with $n \leq 1$. Melott's 
explanation for such a fast clustering growth is as follows: stable clustering 
is not an upper limit to the growth of correlations; whenever the 
initial conditions contain non--vanishing large--scale power, merging makes 
new clusters form, their central density increases with time, which in turn 
enhances the growth of correlations. 
Using N--body simulations in the framework of 
the cold dark matter (CDM) model, Efstathiou (1995) has shown that 
faint blue galaxies can have lower correlation 
amplitude than normal luminous galaxies observed at the present day, as
displayed by the observed value of $\epsilon$, provided that such faint 
galaxies are a transient population associated with dark matter haloes of low 
mass (less than $\sim 10^{12} M_\odot$) and rotation speed. 
A numerical study of the evolution of the 
two--point function both for the matter and halo population has been recently 
carried out by Col\'{\i}n, Carlberg \& Couchman (1996); they obtain a 
scale--dependent $\epsilon$ parameter which is about $1$ for mass particles 
in an Einstein--de Sitter universe, and lower for low--density models. 
A broad range of values (ranging from $-0.2$ to $1$ in the flat case and
reaching lower values in the open case) is obtained for haloes, 
depending on their mean density (see also Brainerd \& Villumsen 1994). 

Recently, Jain (1996) has discussed the reliability of the general 
relation of equation 
(\ref{eq:theor}) in the context of various models. His conclusions are that 
the above parameterization for the evolution of clustering is inaccurate in 
CDM--like models, for two reasons. First, because the 
growth of $\xi(r,z)$ with time on intermediate scales is much faster than the 
$(1+z)^{-3}$ law prescribed by stable clustering at fixed proper separation
(see also Peacock \& Dodds 1996), second, because the boundary between the 
linear, mildly non--linear and stable clustering regimes, occurs at scales 
which rapidly change with time. 

A major advance in this field was represented by the work 
of Hamilton et al. (1991), which first provided a sort of semi--analytic 
model (see also Nityananda \& Padmanabhan 1994), able 
to interpolate between the 
very small--scale behaviour, accurately described by stable clustering, and 
the very large--scale (and/or early--time) one, which is expected to follow
the simple prescriptions of linear perturbation theory. Of course, there also 
exist some more physically motivated models for non--linear gravitational 
clustering, such as the Zel'dovich approximation (Zel'dovich 1970) or 
alternative algorithms (e.g. Sahni \& Coles 1995 for a recent review), like 
the adhesion (Gurbatov, Saichev \& Shandarin 1989), frozen--flow 
(Matarrese et al. 1992) and frozen--potential (Brainerd, Scherrer \& 
Villumsen 1993; Bagla \& Padmanabhan 1994) approximations, which allow one 
to follow the time--evolution of the clustering of collisionless matter on a 
wide range of scales and epochs. However, in contrast to
the Hamilton et al. (1991) ansatz and later modifications of it 
(Peacock \& Dodds 1994; Jain, Mo \& White 1995; Peacock \& 
Dodds 1996), none of these alternative 
algorithms exactly reproduces the correct  very small--scale 
(i.e. strongly non-linear) behaviour.  

An approach similar to the one by Hamilton  et al. (1991) has been applied by 
Peacock \& Dodds (1994, 1996) in Fourier space to evolve the dimensionless 
power--spectrum (actually the contribution to the variance per unit 
$\ln k$) into the non--linear regime. 
These types of approaches have been tested against numerical simulations 
in the linear and mildly non--linear regimes
by Baugh \& Gazta\~naga (1996).  Padmanabhan (1996), 
Bagla \& Padmanabhan (1996), Padmanabhan et al. (1996) and 
Munshi \& Padmanabhan 
(1996) have investigated the more strongly non--linear regime, the latter
authors in particular questioning the applicability of the stable
clustering limit. Sheth \& Jain (1996) suggest that the limit of stable clustering 
on small scales is only expected to ensue on scales where the density 
contrast exceeds $\sim 300-600$. This issue remains controversial.

Besides the problems connected to the non--linear evolution of the clustering 
of the dark matter, one also has to face further non--trivial problems, such as 
the correct definition of the linear bias factor relating -- in the simplest 
possible case -- the object number--density fluctuations to the mass--density 
fluctuation field, and its redshift dependence. Although the idea that
cosmic objects of different types might be biased is ubiquitous in
modern cosmology, the models that have so far been constructed are
relatively crude. This question is therefore one of the major 
stumbling blocks to further progress in the analysis of cosmological
structure formation. The problem is particularly crucial here, for
do we need to know not only the form of the bias at the present epoch
but also its redshift--dependence.

One also has to take care to consider exactly how 
the `intrinsic' redshift--dependent
correlation function of any given class of objects is related to the 
`observed' one, i.e. to the statistic one is actually able to extract from 
the data, which usually involves convolving with the redshift distribution of 
objects in the catalogue, accounting for the mask, correcting for 
redshift--space distortions and for the magnification bias 
induced by weak gravitational lensing, 
and finally adopting an appropriate choice for the redshift binning.

The main point we are making is that one needs to be very careful indeed
about  drawing conclusions on the behaviour of the intrinsic correlation 
function of the given objects and on the scenario of galaxy formation from 
the observed clustering of high--redshift objects, because of intervention
of these difficulties. Constructing a theoretical framework for 
modelling these effects is the main task of this work. 
The plan of the paper is as follows. In Section 2 we define the `observed' 
spatial, angular and projected two--point correlation function for 
high--redshift objects and discuss its connection with the underlying mass 
autocorrelation function. In Section 3 we present a theoretical model for the 
redshift evolution of the correlations of the underlying matter distribution.
In Section 4 we introduce a generalized model for the bias of
a class of objects and how the bias might evolve with epoch, discussing
some special cases of this model in some detail. In 
Section 5 we apply our formalism to the clustering of QSOs and galaxies 
at high redshift. Conclusions are drawn in Section 6, where we also
discuss the relationship of this work to alternative analyses of
clustering evolution.

\section{Description of Clustering}

Let $n_{\rm obs}(x \hat \gamma; z,M)$ be the number--density of objects with 
redshift $z$ that an observer placed in the origin
measures in the angular direction specified by the unit vector $\hat 
\gamma$. Here $x = x(z)$ is the comoving radial coordinate 
corresponding to the redshift $z$, which, in a matter dominated universe is
given by Mattig's formula (e.g. Peebles 1980)
\be
x(z) = {2c \over a_0 H_0} 
{\Omega_0 z + (\Omega_0-2) [ - 1 + (\Omega_0 z + 1)^{1/2}] 
\over \Omega_0^2 (1+z)} \equiv \frac{2c A(z)}{a_0H_0 \Omega_0^{2}(1+z)}\;,
\label{eq:x_z}
\ee
$\Omega_0$ being the density parameter and $H_0$ the present value of 
Hubble's constant. Note that, while in the Einstein--de Sitter case $a_0$ is an
arbitrary length--scale which can be set to unity, in the non--flat case
it is given by
\be
a_0 = \frac{c}{H_0} \vert 1 - \Omega_0 \vert^{-1/2} \;.
\ee
In order to convert dimensionless comoving coordinates to physical comoving ones, 
i.e. expressed in Mpc, one should always multiply by $a_0$; 
so, for instance, the comoving radial 
distance of an object with radial coordinate $x(z)$ is $r(z) \equiv a_0 x(z)$.
The inverse relation of (3) is also useful: 
\be
1 + z(x) = 2 {2y (1-\Omega_0) + \Omega_0 + (2- \Omega_0) 
[1 + y^2(1-\Omega_0)]^{1/2} \over (2-y\Omega_0)^2 } \;, \ \ \ \ \ \ \ \ \
y\equiv {H_0 a_0x \over c} \;. 
\ee
In adopting the above relations we are implicitly assuming that redshift 
distortions are negligible, which is a good approximation in 
dealing with high redshift objects. 
The object number--density will generally also depend on other quantities such 
as mass, luminosity or absolute magnitude in a given wave--band, etc., which 
we generally indicated by the variable $M$; $n_{\rm obs}(x \hat \gamma; z,M)$ 
will be taken to represent the observed number--density of objects per unit 
logarithmic interval of $M$. In the particular models discussed in Sec. 4,
however, we shall make the assumption that relevant physical parameters of the
object are tightly correlated with the halo mass and, in that context, we
shall simply take $M$ to mean the mass of the parent halo.

The probability per unit solid angle to observe an object of type $M$ with 
redshift in the interval $z,z+dz$ is 
\be
\int_{4 \pi} {d \Omega_\gamma \over 4 \pi} 
n_{\rm obs}(x \hat \gamma; z,M) g(z) dz, 
\ee
where 
\begin{eqnarray}
g(z) & \equiv & a_0^3 (1+z)^{-3} \left[1 - \frac{(H_0a_0x)^2}{c^{2}}
(\Omega_0-1)\right]^{-1/2} 
x^2 {dx \over d z} \nonumber\\
 & = & \frac{8c^{3}A^2(z)}{H_0^3 \Omega_0^4(1+z)^{7}}
\left\{ 1 - \frac{4(\Omega_0-1)}{\Omega_0^{2} (1+z)^2}
A^2(z)\right\}^{-1/2}
\left\{ \Omega_0 + (\Omega_0-2) \left[ 1+ \frac{\Omega_0(1+z)}
{2(1+\Omega_0z)^{1/2}}-(1+\Omega_0z)^{1/2}\right]\right\}.
\label{eq:g_z}
\end{eqnarray}
The joint probability per squared unit solid angle to observe two objects of 
type $M_1$ and $M_2$ at redshifts $z_1$ and $z_2$ in directions separated by 
an angle $\vartheta$ is 
\be 
\int_{4 \pi} {d \Omega_{\gamma_1} \over 4 \pi} 
{d \Omega_{\gamma_2} \over 2 \pi} \delta^D\bigl( 
\hat \gamma_1 \cdot \hat \gamma_2 - \cos\vartheta \bigr)
n_{\rm obs}(x_1 \hat \gamma_1; z_1,M_1) 
n_{\rm obs}(x_2 \hat \gamma_2; z_2,M_2) 
g(z_1) g(z_2) dz_1 dz_2,
\ee
where $\delta^D$ stands for the Dirac delta function. 

In practice one calculates different quantities, namely the mean number of 
objects per unit solid angle with redshift in some range ${\cal Z}$ 
and $M$ in a certain domain ${\cal M}$, 
\be
\bar{N}_\Omega =
\int_{\cal M} d\ln M' \int_{\cal Z} d z'
g(z') \int_{4 \pi} {d \Omega_\gamma \over 4 \pi}
n_{\rm obs}( x' \hat \gamma; z',M') \; ,
\ee
and the mean number of pairs at fixed comoving separation $r$ in the same 
redshift interval and $M$ domain,
\ba
\bar N_{pairs}(r) \propto \int_{\cal M} d\ln M_1 d\ln M_2 
\int_{\cal Z} d z_1 d z_2 g(z_1) g(z_2) 
\int_{4 \pi} {d \Omega_{\gamma_1} \over 4 \pi}
{d \Omega_{\gamma_2} \over 2 \pi} \delta^D\bigl( \hat \gamma_1 
\cdot \hat \gamma_2 - \cos\vartheta(x_1,x_2,r) \bigr) \\
\nonumber
n_{\rm obs}(x_1 \hat \gamma_1; z_1,M_1)
n_{\rm obs}(x_2 \hat \gamma_2; z_2,M_2) \;,
\ea
where $\vartheta(x_1,x_2,r)$ is the angle between two sides of a 
triangle whose radial coordinates are $x_1$ and $x_2$, given that the third 
side has size $r\equiv a_0 x_{12}$, as measured from the point with
coordinate $x_1$. In a flat 
geometry, 
\be
\cos\vartheta(x_1,x_2,r) = 
\frac{x_1^2 + x_2^2 - x_{12}^2}{2 x_1 x_2},
\ee
while in the most general non--flat universe case the relation must be 
obtained by inverting the expressions reported below. 
The observed number of objects will generally differ from the real one 
by the catalogue selection function $\phi(z,M)$, which, for
simplicity, we assume to be isotropic,
$n_{\rm obs}(x \hat \gamma; z,M) = \phi(z,M) 
n(x \hat \gamma; z,M)$, where $n(x \hat \gamma; z,M)$ is the `real'
number density of objects.
Once the selection function is known, the mean number of pairs in the sample 
can be compared with different realizations of a theoretical model, 
corresponding to different observers. In practice, however, it is easier to 
compare the 
observed number of pairs directly with the ensemble average of the same 
quantity in the theory. Angular brackets will denote ensemble averages (or 
averages over different spatial locations of the observer). The ensemble 
averaged mean number--density of objects of type $M$ and redshift $z$ will then 
be $\bar n_{\rm obs}(z,M) = \phi(z,M) \langle \bar n(z,M)\rangle$. The two 
averages commute, but the ensemble averaged number--density is direction 
independent, so the further angular mean becomes unnecessary, 
$\bar n_{\rm obs}(z,M) = \phi(z,M) \langle n({\bf x}; z,M) \rangle \equiv 
\phi(z,M) \bar n(z,M)$. 
Similarly, the ensemble averaged mean number of pairs with comoving separation 
$r$, $\langle \bar N_{\rm pairs}(r) \rangle$, 
depends on the cross--correlation $\phi(z_1,M_1) \phi(z_2,M_2) 
\langle n(x_1 \hat 
\gamma_1; z_1,M_1) n(x_2 \hat \gamma_2; z_2,M_2) \rangle$:
from the statistical homogeneity and isotropy of the random field 
$n({\bf x};z,M)$, it follows that the quantity in angular brackets 
is unaffected by quasi--translations (e.g. Weinberg 1972, pg. 
413) and rotations, so that its only spatial dependence can be on 
the separation $R_{12}$ of the point $(x_2, \hat \gamma_2)$ as measured from 
the point $(x_1, \hat \gamma_1)$ (e.g. Osmer 1981; Shanks \& Boyle 
1994\footnote{The formulae given by Shanks \& Boyle 
(1994) contain two errors: 
first, their comoving radial coordinates are incorrectly normalised, 
which leads to wrong determinations of the object separations for any 
$\Omega_0\neq0,~1$; second, $D$ is incorrectly defined as the square root of the true one.}), 
namely 
\be
R_{12} = a_0 \bigl\{D^2 x_1^2 + x_2^2 - 2 D x_1 x_2 \hat \gamma_1 \cdot 
\hat \gamma_2 \bigr\}^{1/2} \;, 
\label{eq:r12}
\ee 
where 
\be 
D = \left[1 - y_1^2(\Omega_0-1)\right]^{1/2} + {x_1 \over x_2} 
\left[1- \left(1- y_2^2(\Omega_0-1)\right)^{1/2}\right] \hat 
\gamma_1 \cdot 
\hat \gamma_2 \;,
\ee
where $y_i=H_0a_0x_i/c$. Equation (\ref{eq:r12}) 
reduces to the usual 
cosine rule, $R_{12}=a_0 \sqrt{x_1^2 + x_2^2 - 2 x_1 x_2 
\hat \gamma_1 \cdot \hat \gamma_2}$ in the 
Einstein--de Sitter case. Notice that the formula (12)
for $R_{12}$ is not symmetric in $x_1$ and $x_2$. This is, however,
only important if $D$ differs significantly
from unity which effectively means that one is dealing with distances of 
order the curvature
radius. Since in any case we expect correlations from objects
at such large separations to be very weak (Stebbins \& Caldwell 1995)
it is possible to ignore
this effect for reasonable cosmological models.
One can then replace the above 
cross--correlation into the expression for $\langle \bar N_{\rm pairs}(r) 
\rangle$ and perform the angular integrations; the effect of the Dirac 
delta function will then be that of fixing $R_{12}=r$. By normalising to the 
expected number of Poisson pairs in the same sample, one obtains (an 
ensemble average estimate for) the correlation function actually measured in 
the sample, namely
\be
\xi_{\rm obs}(r) = N^{-2} 
\int_{\cal Z} d z_1 d z_2 \int_{\cal M} d\ln M_1 d\ln M_2 
~{\cal N}(z_1, M_1) ~{\cal N}(z_2, M_2) \langle \delta_n ({\bf x}_1; z_1,M_1) 
\delta_n ({\bf x}_2;z_2,M_2) \rangle \; ,
\ee
where $\delta_n ({\bf x}; z,M) \equiv [n({\bf x}; z,M) - \bar n(z,M)]/ 
\bar n(z,M)$, while ${\bf x}_2$ is any point whose separation from ${\bf x}_1$ 
is $r$. We also introduced the quantity 
${\cal N}(z,M) \equiv 4 \pi g(z) \phi(z,M) \bar n(z,M)$, representing the 
number of objects in the sample with intrinsic property $M$ in the range 
$(\ln M,~\ln M+d \ln M)$ and redshift in 
the range $(z,~z+dz)$, and its integrals ${\cal N}(z) \equiv 
\int_{\cal M} d\ln M' ~{\cal N}(z,M')$ and 
$N \equiv \int_{\cal Z} d z' {\cal N}(z')$, with
obvious physical interpretation. 
Given our assumptions, negligible redshift distortions and an isotropic 
selection function, the latter formula is exact and general. 
It shows that, in 
principle, one should know the cross--correlation of the random field 
$n({\bf x}; z,M)$ at two different $z$ and $M$. In order to proceed, however, 
we need to make some approximations. First of all we will assume 
separation of property $M$ and position in the form of a linear biasing factor 
relating the object number--density fluctuation $\delta_n$ to the 
mass--density fluctuation $\delta_m$, namely $\delta_n({\bf x};z,M) \approx 
b(M,z) \delta_m({\bf x};z)$, which is a reasonable approximation as long 
as a calculation of the two--point function is concerned. In such a case we 
find
\be
\xi_{\rm obs}(r) = N^{-2} 
\int_{\cal Z} d z_1 d z_2 ~{\cal N}(z_1) ~{\cal N}(z_2)
~b_{\rm eff}(z_1) ~b_{\rm eff}(z_2) \langle \delta_m ({\bf x}_1; z_1) 
\delta_m ({\bf x}_2;z_2) \rangle \; ,
\label{eq:xiobs1}
\ee
where, again, ${\bf x}_2$ is any point whose distance from ${\bf x}_1$ 
is $r$, and we introduced the {\em effective} bias factor 
\be 
b_{\rm eff}(z) \equiv {\cal N}(z)^{-1} \int_{\cal M} d\ln M' ~b(M,z) 
~{\cal N}(z,M') = { \int_{\cal M} d\ln M' ~b(M,z) ~\phi(z,M') ~{\bar n}(z,M') 
\over \int_{\cal M} d\ln M' ~\phi(z,M') ~{\bar n}(z,M')} 
\; .
\label{eq:b_eff}
\ee
In fact, in a wide class of biasing scenarios one expects the bias parameter
to be a monotonically decreasing function of scale (Coles 1993), 
so that the shape of the correlation function of objects at small scales 
is expected to be steeper than that of the mass. However, for simplicity
we shall neglect this point in the subsequent analysis and assume that the 
bias can be described by a constant multiplier of the matter correlation function.
We discuss the importance of bias in the study of clustering
evolution at length in Section 4.

Next, we make the approximation that the dominant contribution to the integral 
in equation (\ref{eq:xiobs1}) comes from points whose redshifts are 
nearly the same. 
As far as the integral is concerned, we can then make the replacement 
$\langle \delta_m ({\bf x}_1; z_1) \delta_m ({\bf x}_2;z_2) \rangle 
\longrightarrow \xi(r, z_{\rm ave})$, where $\xi(r, z_{\rm ave})$ is the mass 
autocorrelation function at some intermediate redshift $z_{\rm ave}$ which we 
can take as $z_{\rm ave} = \bar z \equiv (z_1+z_2)/2$.\footnote{Estimating 
$z_{\rm ave}$ through $z(\bar x)$ instead of $\bar z$ would imply a negligible 
correction in the final integral (actually less than a few percent for 
separations larger than one Mpc, in the model considered below).} 
A rough criterion for 
the validity of this approximation is $\vert \xi''(r,z) / \xi(r,z) \vert 
\Delta z^2 \ll 1$, having indicated by a prime the differentiation with 
respect to $z$, where $\Delta z$ is the width of the given redshift range;
this point is discussed in more detail by Porciani (1996, in preparation). 
At this level the possible spatial curvature of the universe 
can be assumed to affect $\xi(r,z)$ only through its time--evolution, 
since the comoving 
separation of any correlated pairs at the same redshift is certainly much 
smaller than the curvature length. We can then write 
\be
\xi(r, z) = {1 \over 2 \pi^2} \int_0^\infty dk k^2 P(k, z) 
j_0(kr) \; ,
\ee
with $P(k, z)$ the mass fluctuation power--spectrum at $z$. 
The symbol $j_\ell$ will be generally used to denote the spherical Bessel 
functions of order $\ell$. We then arrive to the simple expression
\be
\xi_{\rm obs}(r) = N^{-2} 
\int_{\cal Z} d z_1 dz_2 
~{\cal N}(z_1) ~{\cal N}(z_2) ~b_{\rm eff}(z_1) ~b_{\rm eff}(z_2)
~\xi(r,\bar z) \;. 
\label{eq:xifund}
\ee
The observed correlation function in a given redshift interval is a 
suitable weighting of the mass autocorrelation function with the mean number 
of objects and effective bias factor in that range. Only in the limiting case 
where the objects belong to a narrow redshift interval one is allowed to 
approximate $\xi_{\rm obs}$ by the linear relation $\xi_{\rm obs}(r,z) \approx 
b_{\rm eff}^2(z) \xi(r,z)$, where $z$ is the median redshift of the sample.

Using similar reasoning we can write a simple expression for the
observed angular correlation function in a certain redshift interval and
$M$ domain, in terms of the mass autocorrelation function. 
The result is an extension of the relativistic Limber's formula 
(e.g. Peebles 1980, Sect. 56), accounting for 
a linear bias factor in the relation between $\delta_n$ and $\delta_m$. 
One gets
\be 
\omega_{\rm obs}(\vartheta) = N^{-2} 
\int_{\cal Z} d z_1 d z_2 
~{\cal N}(z_1) ~{\cal N}(z_2) ~b_{\rm eff}(z_1) ~b_{\rm eff}(z_2) 
\xi(r_{12}, z_{\rm ave}) \;, 
\ee
where, neglecting the curvature corrections, 
$r_{12} = a_0 \sqrt{x^2(z_1) + x^2(z_2) - 2 x(z_1)x(z_2) \cos \vartheta}$, with 
$x(z)$ given by Mattig's formula, equation (\ref{eq:x_z}). 
Adopting as usual the {\em small--angle} approximation, one can easily obtain
the handier form 
\be
\omega_{\rm obs}(\vartheta) = N^{-2} 
\int_{\cal Z} d z ~G(z) ~{\cal N}^2(z) ~b^2_{\rm eff}(z) \int_{-\infty}^\infty
d u ~\xi(r(u,\vartheta,z) , z) \;, 
\ee
where $r(u,\vartheta,z) \equiv a_0 \sqrt{u^2 + x^2(z) \vartheta^2}$, with
$x(z)$ given once again by equation (\ref{eq:x_z}), and 
\be 
G(z) \equiv \biggl({d x \over d z}\biggr)^{-1} = {a_0 H_0\Omega_0^2 (1+z)^2 
\over 2c} \left\{ \Omega_0 + (\Omega_0 - 2) \left[ 1 + 
\frac{\Omega_0 (1+z)}{ 2 (\Omega_0z+1)^{1/2}} - (\Omega_0z+1)^{1/2}
\right] \right\}^{-1} \;. 
\ee
Here the approximation of replacing the two redshift dependence of the 
mass autocorrelation function by a single intermediate redshift 
$z_{\rm ave}$ corresponds to the standard procedure (e.g. Peebles 
1980, p. 215). 
 
Another useful quantity to consider is the {\em projected} real--space 
correlation function for objects in the redshift band ${\cal Z}$. 
This statistic can be easily obtained by applying directly the Davis \& 
Peebles (1983) technique to our $\xi_{\rm obs}(r)$ above. 
Defining $r_p$ as the 
separation of a pair perpendicular to the line of sight, 
$r_p=(r_1 + r_2) \tan(\vartheta_{12}/2)$, with $r_1$ and $r_2$ the 
comoving radial coordinates of the members and $\vartheta_{12}$ their angular 
separation, one simply has 
\be
w_{\rm obs}(r_p) 
= 2 \int_0^\infty d y ~\xi_{\rm obs}(\sqrt{r_p^2+y^2}) =
2 \int_{r_p}^\infty d r ~r ~(r^2 - r_p^2)^{-1/2}
~\xi_{\rm obs}(r) \;. 
\ee
In deriving this formula we have neglected peculiar motions, which, 
however, mostly affect separations along the line of sight, and 
assumed Euclidean geometry, which is a reasonable approximation as long as 
the underlying cross--correlation function is evaluated at a 
common mean redshift, as in equation (\ref{eq:xifund}). 

As recently pointed out by Villumsen (1995), if the redshift distribution of 
faint galaxies is estimated by applying an apparent magnitude selection
criterion, then there is a possibility that  a 
magnification bias due to weak gravitational lensing (Turner 1980) 
would intervene in  the relation between the intrinsic spatial correlation function and the 
observed angular one. A similar problem could also affect the 
observed spatial correlation function 
of high redshift objects. This effect has not been taken into account in this
work but we note that it would tend to increase the observed correlation
function of high-redshift objects by adding a source of extra apparent
correlations over and above that produced by intrinsic spatial correlations
of the objects themselves.
 
\section{Modelling the evolution of matter correlations} 

In the light of the preceding discussion it is clear that a theory
capable of predicting the observed correlation function of some class of
objects must first incorporate 
a model for the initial linear autocorrelation function of
the primordial density fluctuations and also a model describing how these
correlations evolve with time. Only after this has been constructed
does it make sense to worry about how a given class of objects relates
to the mass distribution. We defer discussion of this second task to
section 4; in this section we concentrate exclusively on the evolution of
matter correlations.

As an illustrative example to which we shall apply
our method, we assume the  phenomenological 
linear power--spectrum obtained by Peacock \& Dodds (1996) by fitting 
correlation results coming from different samples of galaxies and clusters.
In their analysis they consider a linear power--spectrum of the form
\be
P_{\rm lin}(k,0)=P_0 k^n T^2(k)\ ,
\ee
with the CDM transfer function, as given by Bardeen et al. 
(1986), 
\be
{T(k)}={{\ln(1+2.34q)}\over {2.34q}} \left[ 1+3.89q+(16.1q)^2+(5.46q)^3+
(6.71q)^4\right]^{-1/4}\ ,
\ee
and $q\equiv (k/h {\rm Mpc}^{-1})/\Gamma_{\rm eff}$, where $h$, as usual,
is the Hubble constant in units of 100 km s$^{-1}$ Mpc$^{-1}$. 
Peacock \& Dodds (1996) find the following good fit for the effective shape parameter $\Gamma_{\rm
eff}=0.255\pm 0.017+0.32(1/n-1)$ originally introduced by Efstathiou, Bond \&
White (1992). In CDM models $\Gamma_{\rm eff}$ is related to the density
parameter $\Omega_0$ and to the baryonic fraction $\Omega_b$ by the relation
(Sugiyama 1995) $\Gamma_{\rm eff}=\Omega_0 h \exp[-\Omega_b(1+1/\Omega_0)]$.
White et al. (1996) have recently discussed critical--density CDM models with
high baryon content, as suggested by recent observations, which would help
to reduce $\Gamma_{\rm eff}$ below the `standard' value of 0.5. 
In the following analysis, however, we take $\Gamma_{\rm eff}=0.25$ as an empirical parameter,
without any relation to a true CDM model, assuming $\Omega_0=1$. 

The normalisation is then fixed consistently with the 4--year COBE
data which give $Q_{rms PS} = 18 \pm 1.6 ~\mu$K (Bennett et al. 1996;
G\'orski et al. 1996), for 
$n=1$, which we assume here, and $T_0=2.728 \pm 0.004~$K (Fixsen et al. 1996). 
This leads to $P_0 \approx 6.74 \times 10^5~(h^{-1}~{\rm Mpc})^4$ and 
$\sigma_8 \equiv \sigma_{\rm lin}(8 ~h^{-1}{\rm Mpc}) = 0.65$, in reasonably good 
agreement with the Peacock \& Dodds (1994) amplitude. A more recent analysis
of the COBE 4-year data by Bunn \& White (1996), who used a Karhunen-Lo\`{e}ve
expansion to produce an unbiased estimate of the normalisation,
implies an increase in the amplitude of the power--spectrum of order
16 per cent for a standard CDM model; but changing the normalisation of
our phenomenological model in this way will not alter the conclusions of this
paper.

The question of how the matter correlations evolve into the non--linear regime
has been the subject of considerable recent research activity.
A large fraction of the literature on the clustering of 
high--redshift objects uses either  the self--similarity relation for stable 
clustering or {\em ad hoc} generalizations of it,
as in equations (\ref{eq:theor}) and (\ref{eq:powlaw}), 
which also assume separation of the spatial (in 
proper coordinates) and redshift dependence (e.g. Peebles 1980).
The self--similarity relation, obtained from equation 
(\ref{eq:powlaw}) with $\epsilon=0$ (e.g. 
Peebles 1980), is expected to be the 
asymptotic state resulting from the evolution of scale--free initial 
conditions, $P_{\rm lin}(k,0) \propto k^n$ ($-3<n<4$), in an Einstein--de Sitter 
universe, and leading to $\gamma=3(3+n)/(5+n)$. For more general initial 
power--spectra, the first relation, 
with $\epsilon=0$, should still be valid in the stable clustering regime. 
However, it is well known that stable clustering describes the evolution of 
the mass autocorrelation function only in the strongly non--linear limit, 
$\xi~\magcir 100$ (e.g. Efstathiou et al. 1988), under the 
hypothesis that the peculiar relative velocity of the pair exactly balances 
the Hubble flow to form a bound configuration. 
This model only applies on very small scales (e.g. Efstathiou et al. 1988). 
On the other extreme, one has the linear theory prediction 
\be
\xi_{\rm lin}(r,z) = D_+^2(z) \xi(r,0)\;,
\ee
where $D_+(z)$ is the growing mode of linear perturbations 
[$D_+(z)=(1+z)^{-1}$ in an Einstein--de Sitter model], 
which only applies to large scales and/or at early times, i.e. when 
$\xi \ll 1$. Scale--free initial conditions linearly evolved in 
an Einstein--de Sitter universe lead to the form of equation 
(\ref{eq:powlaw}), with 
$\epsilon=n+2$ and $\gamma=n+3$. 

What one really needs is a model able to smoothly interpolate among the 
two above regimes, so that it can be safely applied in the mildly non--linear 
regime, i.e. that relevant for most objects at the relevant separations. 
Following Hamilton et al. (1991), various authors have recently 
obtained fits of the N--body results of the type 
\be 
\bar \xi(r,z) = B(n_{\rm eff}) F[\bar \xi_{\rm lin}(r_0,z) / B(n_{\rm eff})]\;, \ \ 
\ \ \ \ 
r_0=[1+\bar \xi(r,z)]^{1/3} r \;,
\ee 
where 
\be 
\bar \xi(r,z) \equiv {3 \over r^3} \int_0^r y^2 \xi(y,z) dy = 
{3 \over 2 \pi^2 r} \int_0^\infty dk k P(k, z) 
j_1(kr) \;, 
\ee
and similarly for $\bar \xi_{\rm lin}$ in terms of the linear power--spectrum, 
$P_{\rm lin}(k,z) = D_+^2(z) P_{\rm lin}(k,0)$. In Hamilton et al. (1991) the 
$B(n_{\rm eff})$ factor was set to unity. 
For the Einstein--de Sitter case, Jain et al. (1995, hereafter JMW)
gave theoretical arguments for $B(n_{\rm eff}) \approx (1+n_{\rm eff}/3)^{0.8}$, 
with $n_{\rm eff}$ the  effective spectral index, which for a general, 
i.e. non--scale--free, initial  power--spectrum is defined as 
\be
n_{\rm eff}(z) = \frac{d \ln P_{\rm lin}(k,z)} 
{d \ln k} \big|_{k=k_{nl}(z)},
\ee
where $k_{nl}^{-1}$ is the radius of a 
top--hat window function at which the {\em rms} linear mass fluctuation is 
unity. The JMW formula, which reads 
\be 
F(X) = { X + 0.45 X^2 - 0.02 X^5 + 0.05 X^6 \over 1 + 0.02 X^3 + 
0.003 X^{9/2}} \;, \ \ \ \ \ \ X = \bar \xi_{\rm lin}(r_0,z)/B(n_{\rm eff}) \;, 
\ee 
fits the N--body results, in the range of scales and redshifts where it is 
testable, with an accuracy better than twenty per cent. The JMW fit seems to
perform  better than the others (Hamilton et al. 1991; Peacock \& Dodds 1994, 1996) 
in the highly non--linear regime, which is relevant if one needs to
predict the amount of clustering at early epochs 
(e.g. Baugh \& Gazta\~naga 1996). The formula (29) differs from that
used by Peacock \& Dodds (1996) slightly in the
asymptotic behaviour  for small separations.

Here the linear prediction is obtained for $X \to 0$, while the stable 
clustering behaviour is asymptotically recovered for $X \to \infty$. 
To obtain the differential correlation function, $\xi(r,z)$, we need to 
differentiate the above relation with respect to $r$. 
We obtain 
\be 
\xi(r,z) = { \bigl [1 + B(n_{\rm eff}) F(X) \bigr] F'(X) \Delta \xi_{\rm lin}(r_0,z) 
\over 1 + B(n_{\rm eff}) F(X) - F'(X) \Delta \xi_{\rm lin}(r_0,z)} 
+ B(n_{\rm eff}) F(X) \;, 
\ee
with $F' \equiv dF/dX$ and 
\be 
\Delta \xi_{\rm lin}(r_0,z) \equiv \xi_{\rm lin}(r_0,z) - \bar \xi_{\rm lin}(r_0,z) 
= {1 \over 2 \pi^2} \int_0^\infty dk k^2 P_{\rm lin} (k, z) 
\bigl[ j_0(kr_0) - (3 / kr_0) j_1(kr_0)\bigr] \;. 
\ee 

In Fig. 1 we show the mass autocorrelation function $\xi(r,z)$ obtained from 
the model above and compare it with the linear theory and stable clustering 
predictions for the same linear power--spectrum. Note that the stable 
clustering model always tends to largely overestimate the amplitude of the 
correlation function on quasi--linear scales. 

The qualitative effect of convolving the intrinsic object correlation 
function $b^2_{\rm eff}(z) \xi(r,z)$ with the redshift distribution ${\cal N}(z)$, 
as prescribed by equation (\ref{eq:xifund}), 
can be that of partially diluting the signal. 
Quantitatively the effect can be estimated by using a very 
simple model: we assume no bias, $b=1$, uniform selection function,
$\phi(z,M) = {\rm constant}$, and constant comoving object number--density, 
${\cal N}(z) \propto (1+z)^3 g(z)$. Integration over a typical redshift 
range $\Delta z = 2 $, centered on a median redshift of $1$, shows 
that the observed correlation function systematically underestimates the 
intrinsic one by about ten per cent. The effect can be higher in case 
the redshift--dependence of either the catalogue selection function, or 
of the intrinsic object number--density, preferentially selects lower redshift 
objects, which are generally less biased. 

An extension of this formalism to the open universe case has been obtained 
for the evolved power--spectrum in terms of the linear one by Peacock \& Dodds 
(1996). The simple idea behind their extension is that if collapse occurs at 
high redshift, when $\Omega$ was very close to unity, then non--linear 
correlations at fixed proper separation
still evolve as $(1+z)^{-3}$ at low redshift, what changes
instead is the linear growth factor which is suppressed by a known 
$\Omega$--dependent factor. We shall defer discussion of clustering
in open universe models to a later paper.

\section{Evolution of bias}

We now turn to the issue of how objects trace the mass distribution, i.e.
to the question of the bias and how it might evolve with epoch. In the
framework we are considering here this basically boils down to the
behaviour of the linear bias parameter $b$, where $b^{2}$ is the
factor by which the two--point correlation function of a given class
of objects exceeds the autocorrelation function of the mass fluctuations.
 The clustering of observable objects
is more difficult to model satisfactorily than that of
their parent haloes, being
sensitive to various cosmological and astrophysical effects
as well as to the mass -- or any other intrinsic property --
dependence of the selection function.
There are various possibilities as to how such a bias might originate, and
different biasing schemes are expected to introduce a bias that depends
on redshift in different ways. In this paper we shall discuss {\em four}
simple models which incorporate evolutionary effects in a relatively
straightforward way. 

The first model we consider is the trivial one in which objects trace
the mass at all epochs or, in terms of the bias parameter, that $b=1$
for all values of $z$. We shall use this model as a kind of reference
standard against which to compare the behaviour of other, more complicated
biasing scenarios. For short, we shall call this the {\em unbiased model}.

The next two models we consider are motivated by a recent
discussion  by Mo \& White (1995) who, working within 
the Press--Schechter (1974) formalism,  obtained an approximate formula
(expected to be valid at large separations) for the linear bias relating the correlation 
function of dark matter haloes of mass $M$ to that of mass--density fluctuations. 
As we mentioned in Sec. 2, we now assume that physical parameters of the
objects forming in haloes of mass $M$ are tightly correlated with $M$, so
that the general `type' parameter we used there ($M$) is now taken to
mean the mass of the parent halo. The assumption that the mass of the
host halo determines the parameters of the object it contains may not be
correct in detail, but it does allow the construction of relatively
simple models that illustrate the way our formalism works.
What emerges from these arguments is an expression for the bias
parameter of dark matter haloes of mass $M$ at redshift
$z$, which formed at redshift $z_f$. The Mo \& White (1985) formula reads 
\be 
\xi_{\rm halo}(r,z,M) = b^2(M,z|z_f) \xi(r,z) \;, 
\ee
with 
\be
b(M,z|z_f) = 1 + {1 \over \delta_f} (\nu_f^2 -1) \; 
\label{eq:b_mz}
\ee
in which $\nu_f=\delta_f/\sigma_{\rm lin} (z, M)$ and $\delta_f=
\delta_c D_+(z)/D_+(z_f)$. Here $\delta_c$ is the critical linear
 overdensity for spherical collapse 
($\delta_c=1.69$ in the Einstein--de Sitter case), 
 and $\sigma_{\rm lin}^2$ is the linear mass--variance at redshift 
$z$ filtered in a sharp--edged  sphere containing the mean mass $M$. 
The expression (33) implies a minimum value for the bias 
factor $b(M,z|z_f) \geq 1 - 1/\delta_c$ ($0.41$ for $\Omega_0=1$).
The most relevant feature of this formula in this context
is that it predicts $b 
\approx 1$ at $\nu_f =1$, which for $\bar n(z, M)$ represented by the 
Press--Schechter  mass--function, 
\be
\bar n(z,M) ~d \ln M = \sqrt{2 \over \pi} {{\bar \varrho_0} (1+z)^3 \over 
M  D_+(z) \sigma_{\rm lin}(0, M) } \bigg| {d \ln \sigma_{\rm lin}(0, M) 
\over d \ln M} \bigg| \exp \biggl[ -{\delta_c^2  \over 
2 D_+^2(z) \sigma^2_{\rm lin}(0, M)}
\biggr] ~d \ln M \;, 
\ee
(with $\bar \varrho_0$ the mean mass density at $z=0$), 
roughly corresponds to the peak mass $M_\star(z_f)$, implicitly defined by 
$\sigma_{\rm lin}(z_f, M_\star)=\delta_c/\sqrt{2}$. 
We are interested not in $b(M,z|z_f)$ but the bias of all haloes of
mass $M$ that exist at an epoch $z$, which we call $b(M,z)$. This
latter quantity is, in principle, obtained by integrating equation (33)
over all $z_f\geq z$ taking into account the relevant {\em survival
probability} of a halo of mass $M$ formed at $z_f$ not being destroyed
by merging by the epoch $z$. It is, however, implicit in the Press--Schechter
approach that the only objects that exist at any epoch $z$ are those which
have just formed by the merging of smaller mass units
(Sahni \& Coles 1995, Sec. 3.4). This assumption may be questionable, but
it is nevertheless  consistent, within the
Press--Schechter formalism, to argue that $b(M,z)=b(M,z|z_f=z)$. Looked
at in this way, the formula (33) leads to the result one would expect
from the so--called 
peak--background split (Efstathiou et al. 1988; Cole \& Kaiser 1989), and
reduces to the standard form $b \approx \delta_c/\sigma^{2}$ in the
limit of high peaks (Kaiser 1984), 
implying a $D_+^{-2}(z)$ (i.e.  $(1+z)^2$ if $\Omega_0=1$) redshift 
dependence in such a limit. 

The task of obtaining the form of $b(M,z)$ for 
objects undergoing dissipative collapse, such as galaxies,
is rather more involved than just finding this function for their
haloes. While haloes may merge rapidly and lose their identity, the same
is not necessarily true for the galaxies they contain. For this reason,
Mo, Jing \& White (1996) prefer to use the formula (33) for pairs of
galaxy clusters with separations $r \geq 10 ~h^{-1}$ Mpc, 
assuming that a cluster can indeed be thought of as a single
large halo, despite the fact that it contains many separate galaxies.
There is therefore some ambiguity in how to translate the formula
(33) to a specific situation of a particular class of objects.

One way to incorporate the above formulae into a model
for the observed clustering of objects is to assume that one can
observe all haloes exceeding a certain cutoff mass $M_{\rm min}$ at any
particular redshift, i.e. $\phi(z,M)=\Theta(M-M_{\rm min})$ at any $z$,
where $\Theta(\cdot)$ is the Heaviside step function.
One then has to weight the correlation function
of haloes of mass $M$ by the appropriate number density.
We can thus model the linear bias at redshift $z$ for haloes of mass $M$ as 
in equation (\ref{eq:b_mz}) and get
the redshift evolution of the effective bias factor $b_{\rm eff}(z)$ through
equation (\ref{eq:b_eff}), by weighting it consistently
with the Press--Schechter
mass--function $\bar n(z, M)$.  A plot of $b_{\rm eff}(z)$ for different choices of the minimum cutoff mass in 
$\bar n(z,M)$ is shown in Fig. 2.  
The $z$--dependence of $b_{\rm eff}$ changes 
with $M_{\rm min}$, which also determines its present value, with 
the constant value $0.41$, obtained for $M_{\rm min} \to 0$. 
(Note that, in principle, $M_{\rm min}$ may depend on the depth
of the catalogue.)
Depending on the mass spectrum considered, the redshift dependence 
of $b^2_{\rm eff}$ can be such as to partially or even completely 
balance the evolution of the mass autocorrelation function.
The behaviour of $b_{\rm eff}$ emerging from such a model
can be well--fitted by using the relation
\be
b_{\rm eff}(z) = 0.41 + [b_{\rm eff}(z=0)-0.41] (1+z)^\beta \;.
\label{eq:bfit}
\ee
The resulting fitting parameters are reported in Table \ref{t:fit} for
our choice of initial matter fluctuation spectrum.
One important feature of this model is that specifying a minimum mass for
the objects determines their effective bias factor at $z=0$. If the
value required does not correspond to the empirically--determined
bias parameter for any known class of objects then one must accept
that such objects are transient and have faded sufficiently since the
observed epoch to be missing from local redshift surveys. 
Alternatively, it may be that observational selection effects 
make objects visible at high redshift, but invisible at small
distances, such as might be the case for extended objects of low surface
brightness. We shall call
this the {\em transient model} in the following discussion.

\begin{table}[tp]
\centering
\caption[]{The fitting parameters of the effective bias $b_{\rm eff}$
by using equation (\ref{eq:bfit}).
Column 1: the minimum cutoff mass (in $h^{-1}~M_\odot$).
Column 2: the value of $b_{\rm eff}$ at $z=0$.
Column 3: the parameter $\beta$.}
\tabcolsep 7pt
\begin{tabular}{lcc} \\ \hline \hline
$M_{\rm min}$ & $b_{\rm eff}(z=0)$ & $\beta$ \\
\hline 
$10^{9}$  & 0.50 & $1.92\pm 0.01$ \\
$10^{10}$ & 0.55 & $1.89\pm 0.01$ \\
$10^{11}$ & 0.67 & $1.85\pm 0.01$ \\
$10^{12}$ & 0.90 & $1.79\pm 0.01$ \\
$10^{13}$ & 1.41 & $1.76\pm 0.02$ \\
$10^{14}$ & 2.67 & $1.78\pm 0.03$ \\
 \hline
\end{tabular}
\label{t:fit}
\end{table}

An alternative scenario  based on these considerations 
is to fix the effective bias parameter at $z=0$ so that
it agrees with a known population of present--day 
objects using $b_0=b_{\rm eff}(0) 
\approx 1/\sigma^{nl}_8 \approx 1.46$ -- where by $\sigma^{nl}$ we mean 
the non--linearly evolved {\em rms} mass fluctuation -- 
and evolving it as 
\be
b_{\rm eff}(z) = b_{-1} + (b_0 - b_{-1}) (1+z)^\beta \;, 
\label{eq:bm1}
\ee
with suitable parameters $b_{-1} \equiv b(z \to -1)$ and $\beta$; fixing
$b_0$ effectively determines $M_{\rm min}$ while the previous calculation 
fixed $M_{\rm min}$ in order to determine $b_0$. 
If one therefore imagines that a given population
of objects at the present epoch (with a `known' value of $b_0$) form
via the merging of lower mass haloes, such as is the
case in hierarchical clustering models, then this model would appear to
be appropriate. For short we shall call this {\em merging model} in the subsequent discussion. 

The crucial difference between the transient model and the merging model
is that the latter requires that one identify the population of objects
one is observing at high $z$ with the progenitors of present--day
bright galaxies while the former does not.

Our final alternative is motivated by different arguments.
According to Nusser \& Davis (1994) and Fry (1996), if galaxies 
form at some characteristic redshift $z_f$ by some non--linear process
which induces a bias parameter at that epoch $b_f$
and if their subsequent motion is purely caused by gravity, then continuity 
leads to a redshift--dependent bias factor which evolves approximately 
as
\be
b(z)= 1+ (b_f-1) {1+z  \over 1+z_f} \;, \ \ \ \ \ \ \ \ \ z<z_f \;,
\ee
for $\Omega_0=1$. The obvious consequence of this
simple model is that the bias factor always tends to unity
as time goes on. 
Allowing for some spread in the values of $b_f$ and $z_f$, 
depending e.g. on the mass of the galaxy, would not 
greatly change this evolution law. We call this model the 
{\em object--conserving model}. Equation (37) is consistent with
the Mo \& White (1995) formula, equation (33), provided that the
effect of merging is negligible.

A linear dependence on $1+z$ as obtained in the object-conserving
model would also be obtained if objects were 
assumed to form at high peaks of the linear density field above some 
threshold $\delta_c= \nu \sigma_{\rm lin}$ threshold, 
with fixed $\nu$ in accordance with
equation (33).  This model has been recently adopted by Croom \& Shanks (1996) 
to  fit the quasar two--point function.

Notice that all four of these models can be thought of as special
cases of the merging model whose behaviour is described by equation
(36). The transient model has its parameters fixed by the choice
of halo mass cutoff. The merging model has parameters determined by the
choice of bias parameter at $z=0$. The object conserving model has
$b_{-1} =\beta=1$, but otherwise free parameters. The case $b=1$
for all epochs can also be accomodated by 
equation (36): a bias of unity in a merging
model would occur if objects preferentially formed in haloes with
mass close to $M_\star$ or if, for some reason, such objects
were selected observationally.

These models are not exhaustive of all the possibilities, but they do
serve to illustrate the spread in possible behaviours expected with
fairly minimal assumptions about the bias. Alternative biasing models, not discussed 
further in this paper, have been considered by, for example, Coles (1993) 
who discusses a general class of 
local bias models and  Catelan et al. (1994), who define a 
{\em weighted bias} algorithm, according to which the biased density field 
coincides with the mass--density whenever the latter exceeds some fixed 
threshold. As mentioned in Section 2, `generic' local bias models
predict that $b$ is a function of scale, a fact we shall ignore in this
analysis. This introduces the possibility that the {\em shape}
as well as the amplitude of the
biasing relation changes with epoch, allowing
for an even wider range of possibilities. We shall defer this question
to future work.

\section{Applications}
In this section we illustrate the use of
our formalism by applying it to a example data sets. 
Given the mass autocorrelation function, each class is characterized by {\em i)} its
redshift distribution ${\cal N}(z)$, which automatically accounts for the
catalogue selection function, and {\em ii)} the effective bias parameter
$b_{\rm eff}(z)$ which we discussed in the previous Section.
 Note that our method is semi--empirical,
in the following sense: the theoretical mass--function is only employed as the
appropriate weighting factor to deduce the effective bias factor, but we never
require that it predicts the observed redshift distribution of objects in the
catalogue, modulo the selection function; in our applications, in fact, we will
take ${\cal N}(z)$ as obtained from the data. 

\subsection{Spatial QSO correlation function} 
Quasars are a class of objects to which we may apply this formalism. 
Various authors have looked at the spatial quasar clustering and at their
redshift evolution (e.g. Shanks et al. 1987; Iovino \& Shaver 1988; Andreani 
\& Cristiani 1992; Andreani et al. 1994; Shanks \& Boyle 1994; 
Croom \& Shanks 1996). 

According to Efstathiou \& Rees (1988), Haehnelt (1993) and 
Katz et al. (1994), every halo of mass greater than some threshold 
$M_{\rm min}  \sim 10^{11} - 10^{12} ~h^{-1} ~M_\odot$ harbours a
QSO with a lifetime $t_Q$ which we can assume to be much less than
the time it takes for the halo to lose its identity by merging.
In this case QSOs existing at a particular epoch sparsely sample
the distribution of haloes with mass greater than $M_{\rm min}$
at that epoch. This implies that $b_{\rm eff}(z)$ for 
quasars can be calculated using the transient model.

The spatial correlation function of quasars in the redshift range 
$0.3 < z < 2.2$ is shown in the upper panel of
Fig. 3; the redshift distribution 
${\cal N}(z)$ for the Durham/AAT survey is taken from Shanks \& Boyle (1994)
while the observational data are the updated estimates 
from Croom \& Shanks (1996). 
The effective bias $b_{\rm eff}(z)$ is calculated from equation (\ref{eq:b_eff}), 
where  $\bar n(z,M)$ is modelled by the Press--Schechter mass--function, 
with cutoff  mass in the range $10^{11} - 10^{12}~h^{-1}~M_\odot$. 
As it is clear from the plot, given the uncertainty on the minimum 
mass of haloes able to host a quasar, our CDM--like model 
accounts for all the observational data, with the (possible)
exception of the  point at $\simeq 5 h^{-1}$ Mpc,  at which the
difference is less than $2\sigma$. 

In the other panels of Fig. 3
we plot the spatial correlation function of the same QSO catalogue
in two different redshift ranges: 
$0.3 < z < 1.4$ (centre) and $1.4 < z < 2.2$ (bottom).
While the results at higher redshifts are in good
agreement with the observational data, the points at $r \sim 10$
and at $r \ge 20 h^{-1}$ Mpc for $0.3 < z < 1.4$ show a $2 \sigma$ deviation.
Notice that the predicted evolution of $\xi$ over these ranges is
rather small, even for the upper band of the shaded region. In fact,
the data at these scales suggest stable clustering in comoving
coordinates. The
possible discrepancies between the model and the data, however,
are on very small scales, where $\xi_{\rm obs}>1$. This suggests 
that a more decisive test of this model could be
constructed by quantifying the clustering of QSOs on scales
smaller than $r_c$. Notice that there is some indication that
a scale--dependent bias is required to reconcile the observed
correlation function with that predicted by our model, but that
this is to some extent predicted by realistic biasing scenarios
(Coles 1993).

\subsection{Angular and projected galaxy correlation functions} 
While the choice of bias model seems relatively straightforward for
QSOs, this is is not the case for galaxies. Various plausible
arguments can be made for choosing any of the models we discussed
in Section 4.

The first possibility we consider is that galaxies represent at {\em any} 
redshift an unbiased sample of the mass distribution, i.e. that $b_{\rm eff}(z)=1$. 
According to the model of equation (\ref{eq:b_mz}), for the bias of dark 
matter haloes this situation would occur if galaxies were preferentially formed 
from haloes with mass close to $M_\star(z_f)$ and survived until the
epoch $z$. 

Another possibility would be that faint galaxies are subunits 
that merge to make up more luminous galaxies (Broadhurst, Ellis \& Glazebrok 
1992; Clements \& Couch 1996; Baugh, Cole \& Frenk 1996), in which case, 
using the redshift dependence of $b_{\rm eff}(z)$ obtained 
in the previous section for dark matter haloes, looks a sensible choice. 
This model too fits into equation (36), for $b_{-1} = 0.41$, 
$b_0=1/\sigma_8^{nl}\simeq 1.46$ and $\beta \approx  1.8$. 

One might also assume that faint galaxies form a transient population 
of dwarf galaxies which has now faded away (Babul \& Rees 1992; 
Lacey et al. 1993). This would motivate the transient model which,
as we have noted, can also be described by equation (36). For illustrative
purposes we incorporate a minimum mass of $M_{\rm min}=10^{11}h^{-1} M_\odot$,
with consequent values of the parameters given in Table 1.

If galaxies form at some particular redshift and then evolve
without losing their identity and thus
simply following the continuity equation thereafter, the
object-conserving model seems to be appropriate. In this picture,
faint galaxies of a particular type would be identified with
`ordinary' luminous galaxies at high redshift (e.g. Tyson 1988). 
It is natural to apply this scenario to, for example, present-day
bright spirals which appear to have survived undisturbed for some
time by interactions or mergers with objects of a similar mass: the
corresponding value of $b_0=1/\sigma_8^{nl}\simeq 1.46$.

Given the plausibility of each of these models, we now compare all
of them with the best currently available data on the evolution of
galaxy clustering with redshift using the model described in Section 2
to evolve the underlying matter correlations.

In the top panel of Fig. 4 we plot the angular correlation of galaxies out to
$z=1.6$ for the Canada--France Redshift Survey (CFRS). The galaxy redshift
distribution is taken from Crampton et al. (1995). The observational data have
been obtained by Hudon \& Lilly (1996) and refer to the same catalogue. These
data are uncorrected for the integral constraint, spurious field to field
variations or for stellar contamination. Two different methods were used
by Hudon \& Lilly to estimate $\omega(\theta)$ from the observational data: 
in the first the
angular correlation is obtained by averaging the different estimates in each of
the 24 fields composing the catalogue (local method, open circles); in the second
the counting of pairs is done over all cells in the sample (global method,
filled squares). The true unbiased angular correlation is bracketed by the two
methods. 

The bottom panel of the same figure shows  our prediction for a sample of 222 galaxies 
with $z \leq 1.6$ selected from the Hawaii 
Keck K--band survey (Cowie et al. 1996). The redshift distribution 
${\cal N}(z)$ and the correlation data are taken from Carlberg et al. (1996). 
In order to take into account 
the dilution produced by the uncorrelated foreground
stars, we correct the original 
data by a factor of $(1-f_*)^{-2}$, where $f_*$ is
the fraction of catalogued objects that are stars. Carlberg et al. (1996)
estimate for their catalogue $f_*=1/4$.

Note that the CFRS data are well described by both the unbiased and
transient  model while the Keck data, with their larger errors,
are compatible with the galaxy-conserving model and, marginally, with
the merging model.

In Fig. 5 we present our prediction for galaxies in the Hubble Deep Field
(HDF). Villumsen, Freudling \& da Costa (1996) fit the 
redshift distribution  by 
\be
{\cal N}(z)=2.723 {z^2\over z_0^3} \exp [-(z/z_0)^{2.5}]\ ,
\label{eq:nz_hdf}
\ee 
where $z_0$ is approximately the median redshift. 
Alternative estimates of the redshift distribution for these galaxies
are presented by Lanzetta, Yahil \& Fern\'andez--Soto (1996), but for
illustrative purposes we adopt the former results here.
We consider four samples with
magnitude limits $R=26,27,28,29$, with corresponding median redshifts
of  $z_0=1.35,1.54,1.71,1.87$, respectively. Observational estimates of the
correlation functions for these
samples have also been obtained by Villumsen, Freudling \& da Costa (1996).
After correcting the results to include the integral constraint, they fitted
their angular correlations by assuming a power law 
\be
\omega(\theta)= A \theta^{-\delta}
\ee
with fixed slope $\delta=0.8$. The shaded region in the plots refers to the
$1\sigma$ range allowed by their fits. 

Again the unbiased model appears to provide the best fit over all the
the samples selected: the other biasing models yield too high an amplitude
for the deeper samples. There is also some indication that the actual
correlation function of galaxies in this catalogue is steeper than
the model predicts, though this may in principle be accounted for
by allowing the bias factor to depend on scale in the expected
fashion (Coles 1993).

In Fig. 6 we plot the projected correlation function for 
galaxies with CFRS redshift distribution and compare it with observational 
data from the same catalogue (Le F\`evre et al. 1996). 
The data are divided in three ranges in redshift:
$0.2 < z < 0.5$, $0.5 < z < 0.75$ and $0.75 < z < 1$.
Note that the observational data, originally plotted as a function of the 
proper projected separation at the median redshift of three considered 
strips ($z=0.34$, $z=0.62$ and $z=0.86$ respectively),
have been rescaled to the comoving, projected 
separation $r_p$. Again the merging model and the number-conserving
model overpredict the correlations, and the overprediction is worse
for the samples at higher $z$. The unbiased and transient models are,
however, good fits to the data.

Fig. 7 shows our prediction for galaxies for the 
Keck sample, with redshift distribution and 
correlation data from Carlberg et al. (1996). 
Two different strips are here considered: $0.3 < z < 0.9$
and $0.9 < z < 1.5$.
Observational data are rescaled  as before at the median redshift, $z=0.6$ 
and $z=1.1$ respectively. Again the merging model seems to be excluded by
the high redshift data; the other three alternatives are consistent.
Again, there is some evidence that, at high redshift, the shape of
the galaxy correlation function is wrong, possibly because of
scale-dependence of $b$.

\section{Conclusions}

The principal result of this analysis is that uncertainties in the
evolution of the bias parameter with epoch $z$ are potentially the
dominant consideration to be taken into account when testing theoretical
models against galaxy clustering evolution data. Looking at this
in another way, one can say that the rough consistency of observed
clustering data at high redshift with unbiased models can place
strong constraints on the bias invoked in any particular theory
of structure formation. One has to be very careful about postulating
a significant bias at $z=0$ without thinking carefully about how the
bias was introduced, as plausible scenarios give a bias parameter
that evolves strongly with cosmic epoch in contrast with
the observations. Since all the bias models
we consider can be described by the same formula (36), this
basically argues that, at least within the class of models considered,
the bias must be small at the present epoch so as not to increase
too drastically with redshift. Notice further that any extra
correlations introduced by the effects of magnification bias
induced by gravitational lensing actually strengthen this conclusion, 
as the observed clustering
must include both the intrinsic clustering of the objects and
the contribution from correlated shear.

In the case of QSOs, we find that a simple model in which
quasars trace the distribution of haloes is roughly consistent with
the clustering data supplied by Croom \& Shanks (1996) for
all epochs. This conclusions contrasts with the claim of
Croom \& Shanks that the apparently low level of clustering
evolution requires either a strong bias or a low density universe
(or both). The important points of disagreement 
with Croom \& Shanks (1996) are that 
({\em i}) their conclusions are based on a
standard CDM model (i.e. $\Gamma_{\rm eff}=0.5$) for the initial
power--spectrum, ({\em ii}) they use linear theory to evolve the matter
fluctuations and ({\em iii}) they adopt a model for the evolution of bias
which resembles our object-conserving model. 

It is pertinent to compare our analysis of galaxy
clustering data with a recent paper by
Peacock (1996). By comparing estimates of the power--spectrum of
clustering at $z=0$ and the CFRS data on redshift evolution with
the expected non--linear evolution of a phenomenological model
for the power--spectrum he concludes that only an unbiased model
for galaxy correlations in a low density universe is compatible
with the observations. We have not studied low-density models in
this paper, so we do not comment on whether such models offer a 
better explanation of the power--spectrum. There are, however,
 some important points of difference between our analysis of Einstein-de Sitter models
and that of Peacock (1996). First, the model for the initial power--spectrum 
adopted by Peacock is of a slighty different form to ours:
he uses a model of two power laws with a transition region in between.
Since our model is not exactly a power law on small scales
there will be some differences in the consequent clustering
behaviour starting from these two different initial assumptions.
These differences might well alter the behaviour of the correlation
function on the relatively small scales where its amplitude is
known accurately. Secondly, Peacock (1996) uses a different fitting
formula for the non--linear evolution of correlations, based on
that of Peacock \& Dodds (1996). Third, Peacock (1996)
does not consider models for the evolution of $b$, preferring instead
to assume that the bias factor has the same value at all redshifts.
Finally, we do not include clustering data obtained from
catalogues at zero redshift in our analysis because 
different surveys may either involve physically different objects or
possess different observational selection functions (or both).

These differences are important when we compare our main
conclusion (that unbiased or mildly biased clustering developing
from our initial spectrum in a universe with $\Omega_0=1$ is
consistent with the clustering data) with Peacock's conclusion 
that he cannot reproduce the observed clustering with his model
in a universe with $\Omega_0=1$ unless he invokes a particular
form of biasing to counteract the small--scale evolution of
the matter power--spectrum. Since the differences in our approaches
manifest themselves most strongly on small scales, these are
probably the origin of our disagreement on this point. 
In any case, we do not attempt to fit the detailed shape of the IRAS/APM
power--spectrum (for reasons mentioned above) and this is his
main argument that a specific form of the bias is necessary if
$\Omega_0=1$. The CFRS and Keck data, in themselves, are
well explained by our model. Nevertheless, it is obviously important
to understand the origin of present--day bright galaxies such as those
constituting the APM survey. It emerges from our analysis that there
is a problem explaining the high value of $b_0\simeq 1.4$ required
by our CDM--like model for the these galaxies. According to the
models  we have presented, a population of present-day objects with
$b_0>1$ should be identifiable with some class of high--redshift objects
with an even higher value of $b$, unless the present--day objects
are contained within haloes having a minimum mass of order $10^{13}M_\odot$
(see Fig. 2) and which is very different from that corresponding to
any observable high--redshift objects. An alternative solution to
this conundrum that does not require  a `special'
population of objects to be identified with present--day galaxies
is to invoke $\Omega_0<1$, because 
then there is no need for bright galaxies at the present epoch
to have  $b_0>1$. A model with
$\Omega_0<1$ and in which bright galaxies trace the mass is
probably consistent with both the high--redshift data and  local measures
of clustering, as argued by Peacock (1996).

The discrepancies between our conclusions for both QSOs and galaxies
and the studies of Croom \& Shanks (1996) and Peacock (1996) respectively,
serve as an important reminder of the model-dependence
of this type of analysis and of some of the residual theoretical
uncertainties. In particular, it is important to understand the small-scale
stable clustering limit in more detail if this technology is to be
improved. Above all, however, we need to develop much better defined models
for the form and evolution of biasing in a general context, for even in 
the simplest case in which the bias is a constant linear multiplier
there is considerable ambiguity in the modelling of its evolution.

\section*{Acknowledgments.}
We would like to thank 
P. Andreani for useful discussions, 
B. Jain for detailed explanations of a number of 
technical points and J. Peacock for his very helpful advice. 
R.G. Carlberg, S.M. Croom, O. Le F\`{e}vre
and T. Shanks are warmly thanked for providing us
electronic versions of their results.
Italian MURST is acknowledged for partial financial support. 
PC is a PPARC Advanced Research Fellow; he would like to thank the Department 
of Astronomy in Padova for their 
hospitality during a visit. SM would like to thank 
QMW for hospitality under the PPARC Visitors grant.

%\clearpage

\newpage

\section*{Figure Captions}

\noindent
{\bf Figure 1.}
The mass autocorrelation function $\xi$ as a function of the comoving 
separation $r$, at different redshifts: $z=3$ (top left), $z=2$ (top right),
$z=1$ (bottom left) and $z=0$ (bottom right). 
Different lines refer to different models for the clustering evolution:
JMW fitting formula (solid line), linear theory (dotted line) and stable 
clustering (dashed line).

\noindent
{\bf Figure 2.}
The effective bias $b_{\rm eff}$ as a function of the redshift $z$. Different lines
refer to different values of the minimum mass in the Press--Schechter 
mass--function, ranging from $10^9$ to
$10^{14} ~h^{-1}~M_\odot$, from bottom to top. 

\noindent
{\bf Figure 3.}
Theoretical prediction for the spatial correlation function of quasars, with
${\cal N}(z)$ given by the Durham/AAT redshift distribution
(Shanks \& Boyle 1994). Different panels
refer to different redshift ranges: whole range $0.3 < z < 2.2$
(top), $0.3 < z < 1.4$ (centre) and $1.4 < z < 2.2$ (bottom). The shaded
region shows the values obtained if the effective bias is obtained with a
minimum mass in the band $M_{min}= 10^{11}$ (lower line) and $10^{12} ~h^{-1}
~M_\odot$ (upper line). The open squares and relative error bars refer to the
Croom \& Shanks (1996) results for the observed QSO two--point function in the
same sample. 

\noindent
{\bf Figure 4.} 
Theoretical prediction for the angular galaxy correlation function. Top 
panel: galaxies with $z < 1.6$ and ${\cal N}(z)$ given by the CFRS redshift 
distribution (Crampton et al. 1995). Correlation data are from Hudon \& Lilly 
(1996) and are obtained by using two different methods which bracket the
true values: the local and global determinations
are shown by open circles and filled squares, respectively.
Bottom panel: galaxies in the redshift range $0 < z < 1.6$ selected 
from the Hawaii Keck K--band survey, with redshift distribution and 
correlation data from Carlberg et al. (1996). 
The original data are corrected to take into account the dilution produced
by the uncorrelated foreground stars.
Different bias models are considered: unbiased model [$b(z)=1$; solid line];
galaxy--conserving model [equation (\ref{eq:bm1}) with $b_0=1.46$,
$b_{-1}=\beta=1$; dotted line]; merging model {equation (\ref{eq:bm1}) with
$b_0=1.46$, $b_{-1}=0.41$ and $\beta=1.8$; short dashed line]; transient
model [equation (\ref{eq:bm1}) with $b_0=0.67$, $b_{-1}=0.41$,
$\beta=1.85$; long dashed line]. The latter model almost
coincides with the unbiased case. 

\noindent
{\bf Figure 5.} 
Theoretical prediction for the angular galaxy correlation function
for the Hubble Deep Field. 
The shaded region in the plots refers to the
$1\sigma$ range allowed by the fit obtained on the observational data
by Villumsen, Freudling \& da Costa (1996). The panels
show the results for four $R$ magnitude limited 
samples with different median redshift $z_0$: 
$R<26$ and $z_0=1.35$ (top left),
$R<27$ and $z_0=1.54$ (top right),
$R<28$ and $z_0=1.71$ (down left)
and $R<29$ and $z_0=1.87$ (down right).
Different bias models are shown as in Fig. 4. 

\noindent
{\bf Figure 6.} 
Theoretical prediction for the projected galaxy correlation function
of the CFRS sample. The redshift distribution is given
by Crampton et al. (1995). 
Correlation data are from Le F\`evre et al. (1996). 
Different panels refer to different strips in redshift:
$0.2 < z < 0.5$ (top),  $0.5 < z < 0.75$ (centre) and
$0.75 < z < 1$ (bottom). 
Different bias models are shown as in Fig. 4. 

\noindent
{\bf Figure 7.} 
Theoretical prediction for the projected galaxy correlation function
of the Hawaii Keck K--band survey. The redshift 
distribution and correlation data are from Carlberg et al. (1996). 
Different panels refer to different strips in redshift:
$0.3 < z < 0.9$ (top) and $0.9 < z < 1.5$ (bottom).
Different bias models are shown as in Fig. 4; the transient model
almost coincides with the $b=1$ (unbiased) model in the upper panel. 
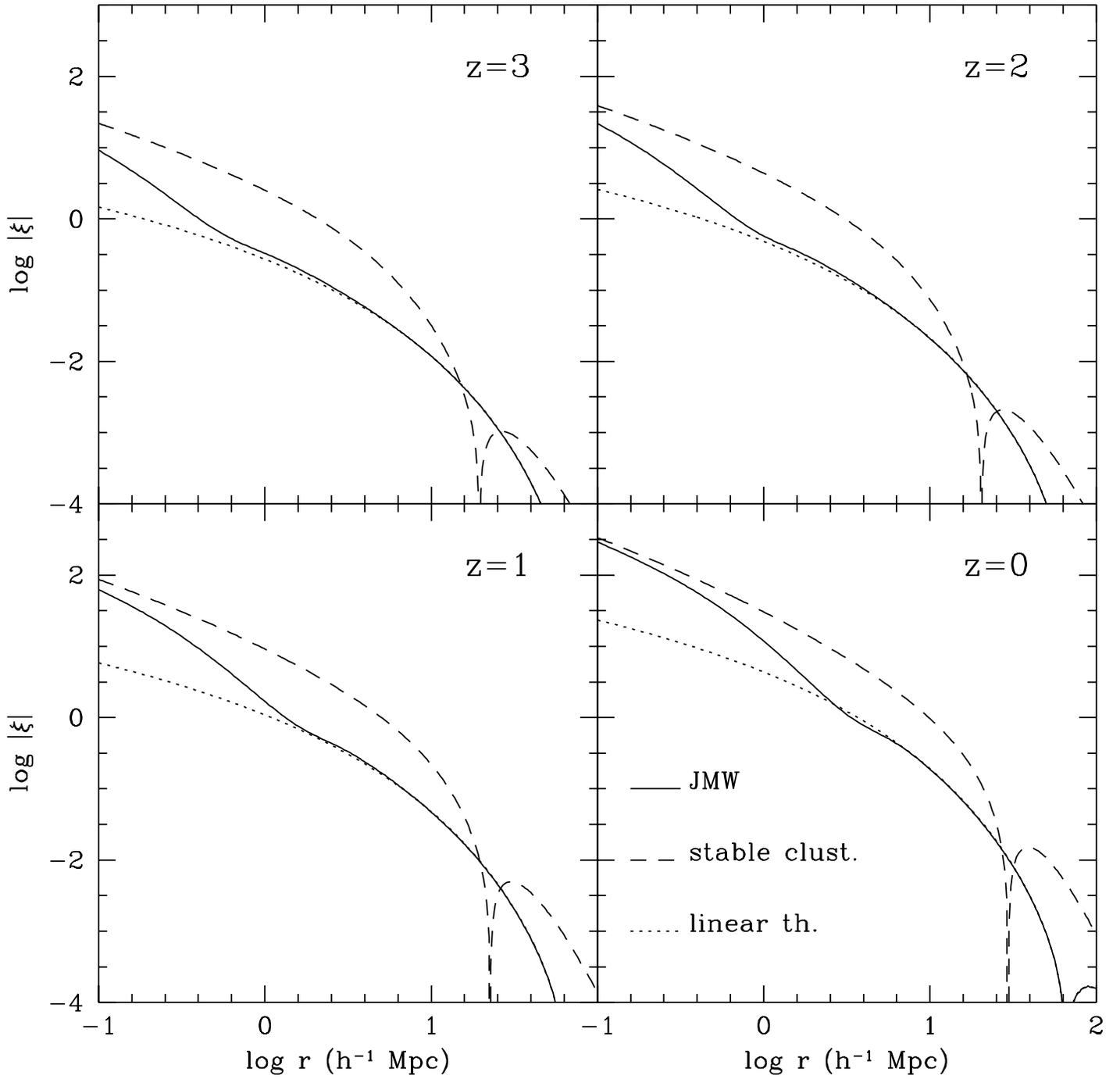
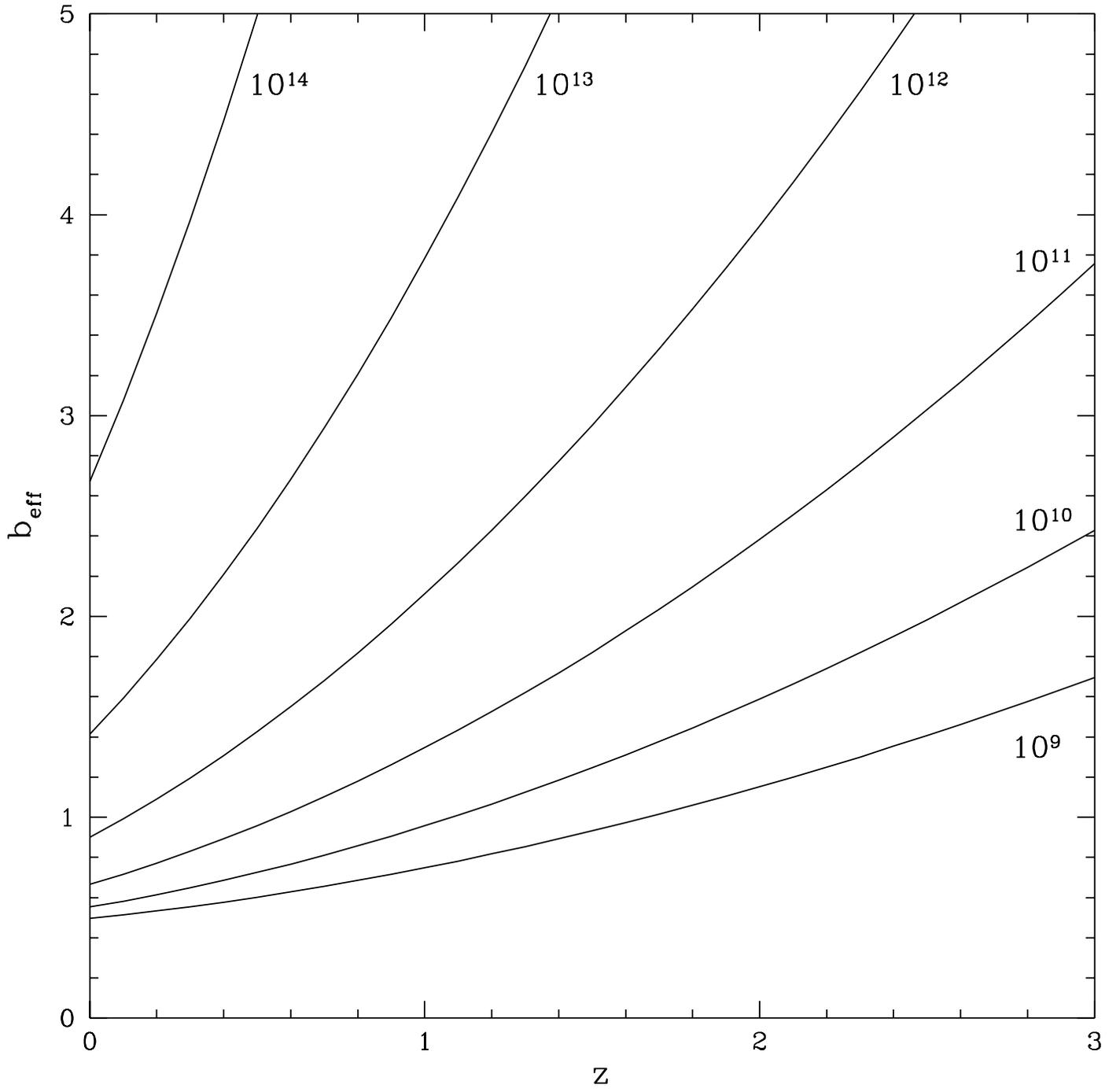
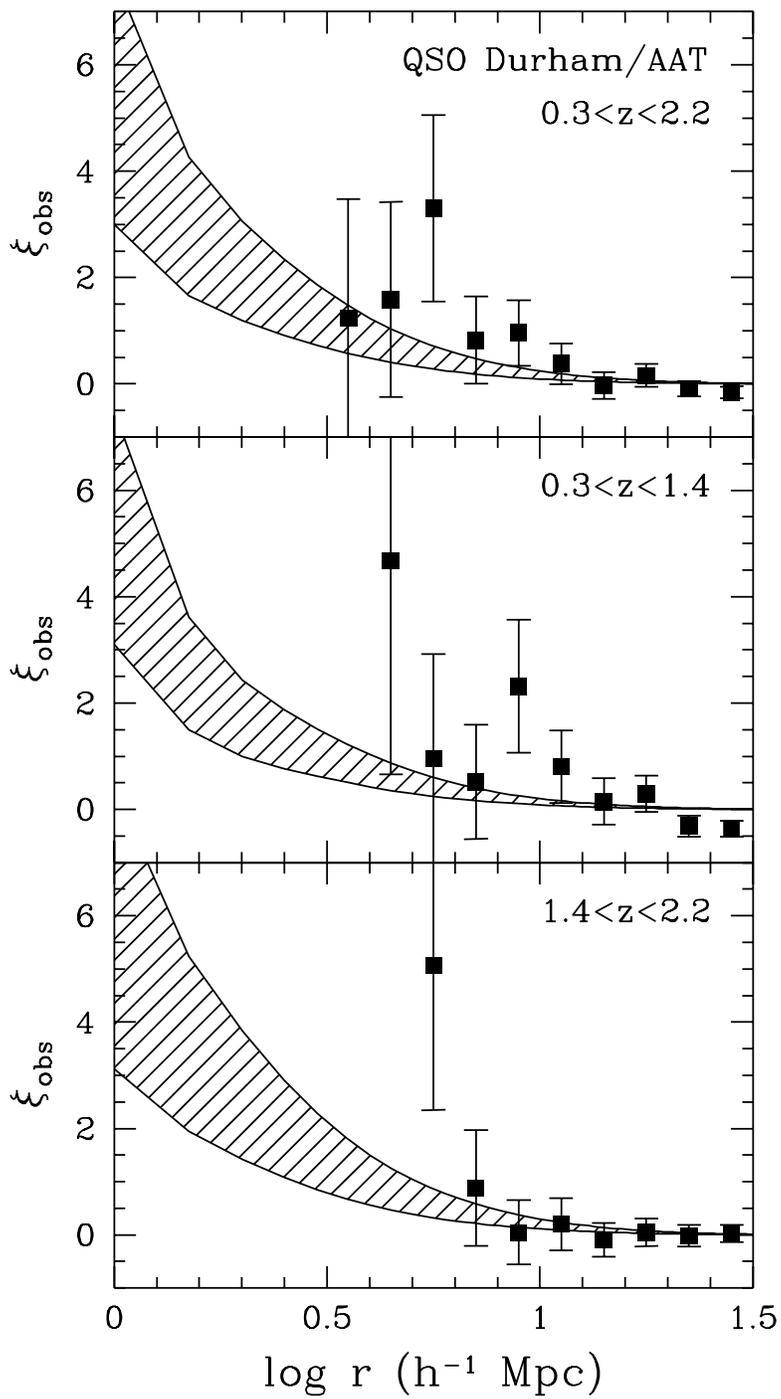
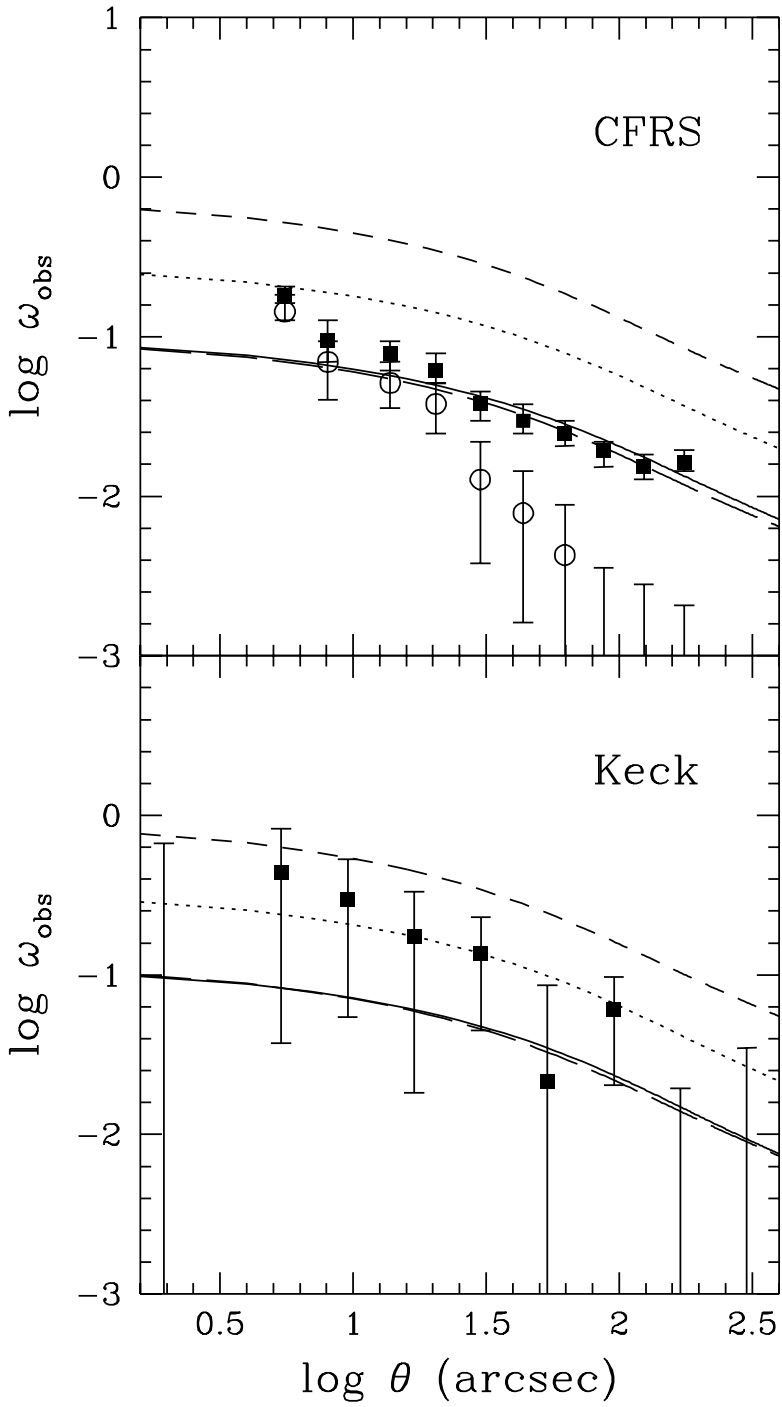
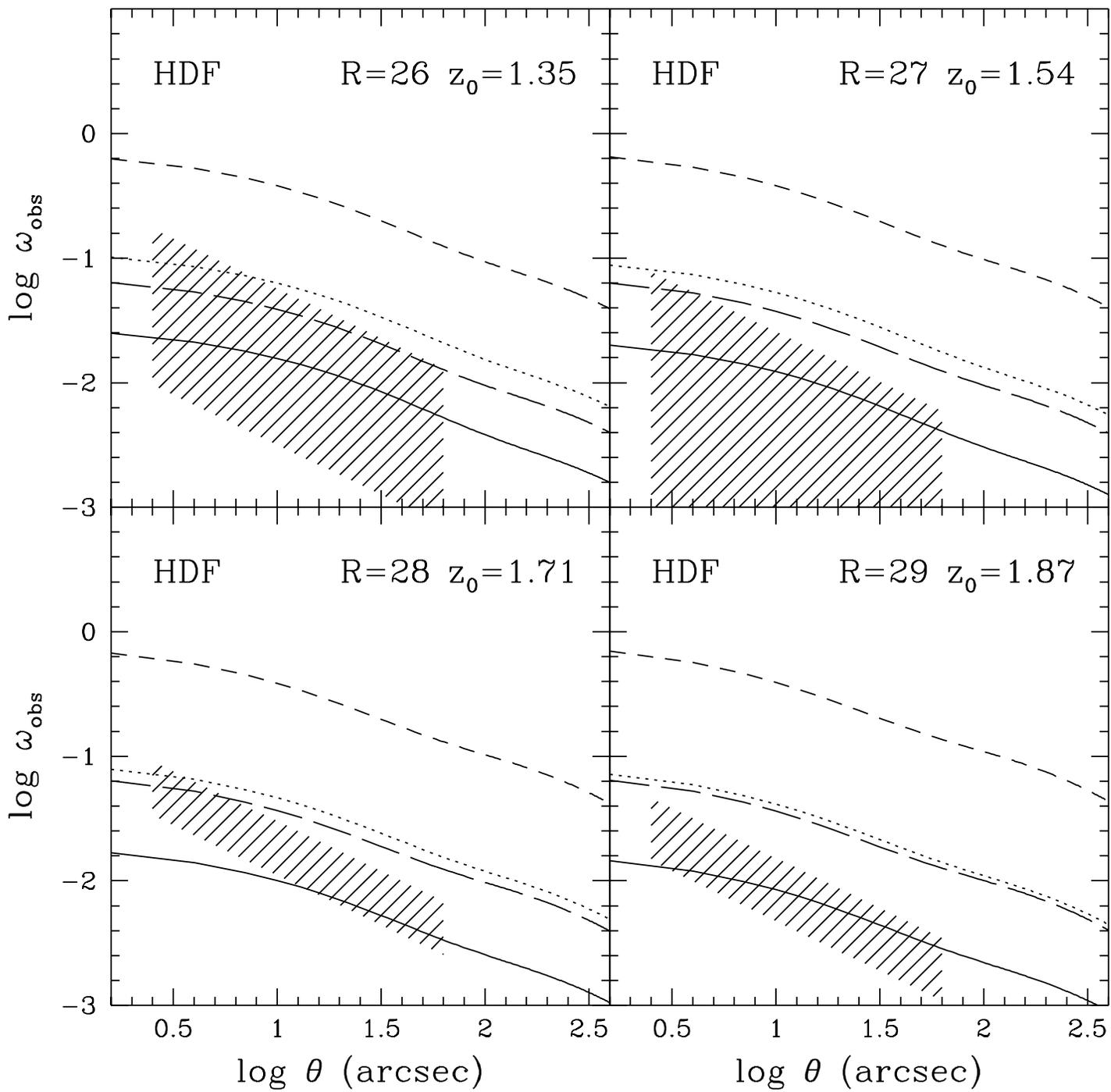
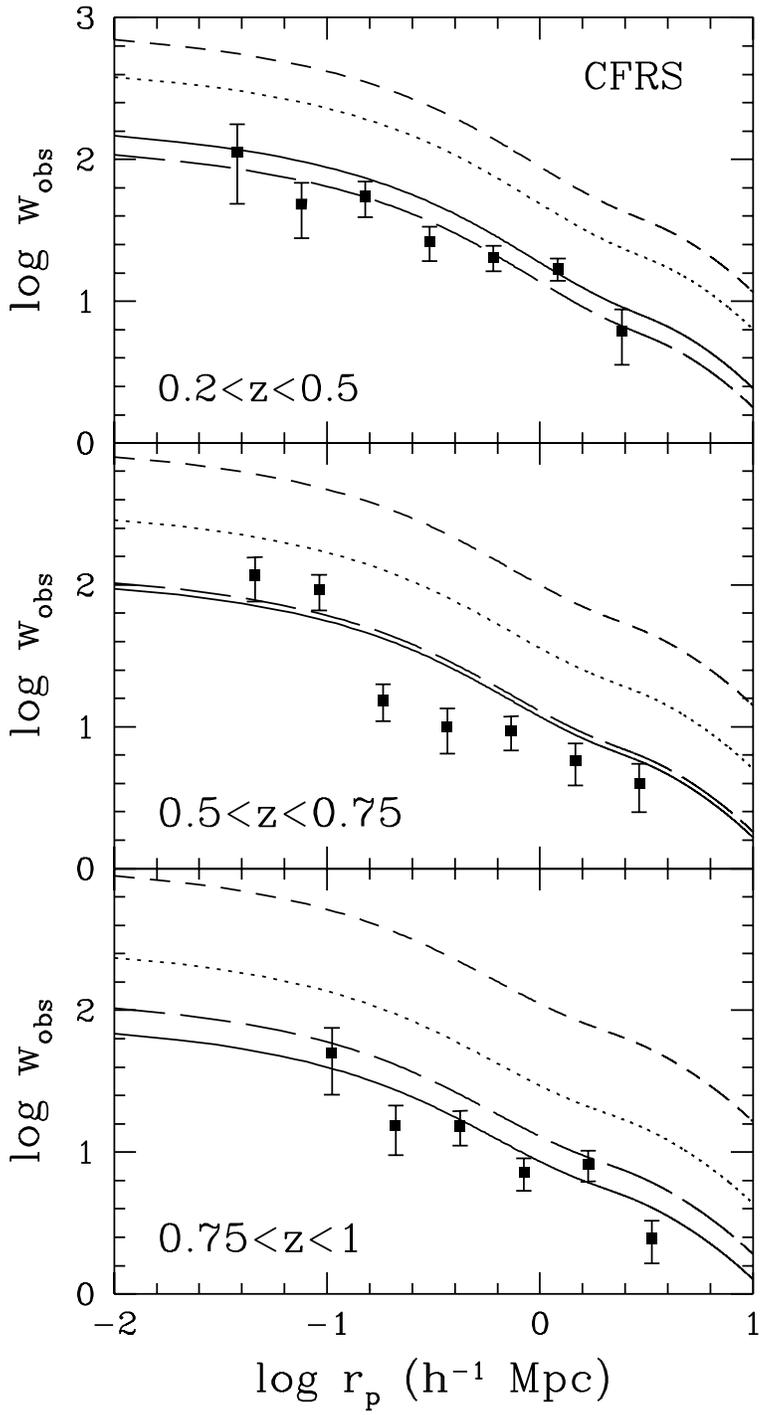
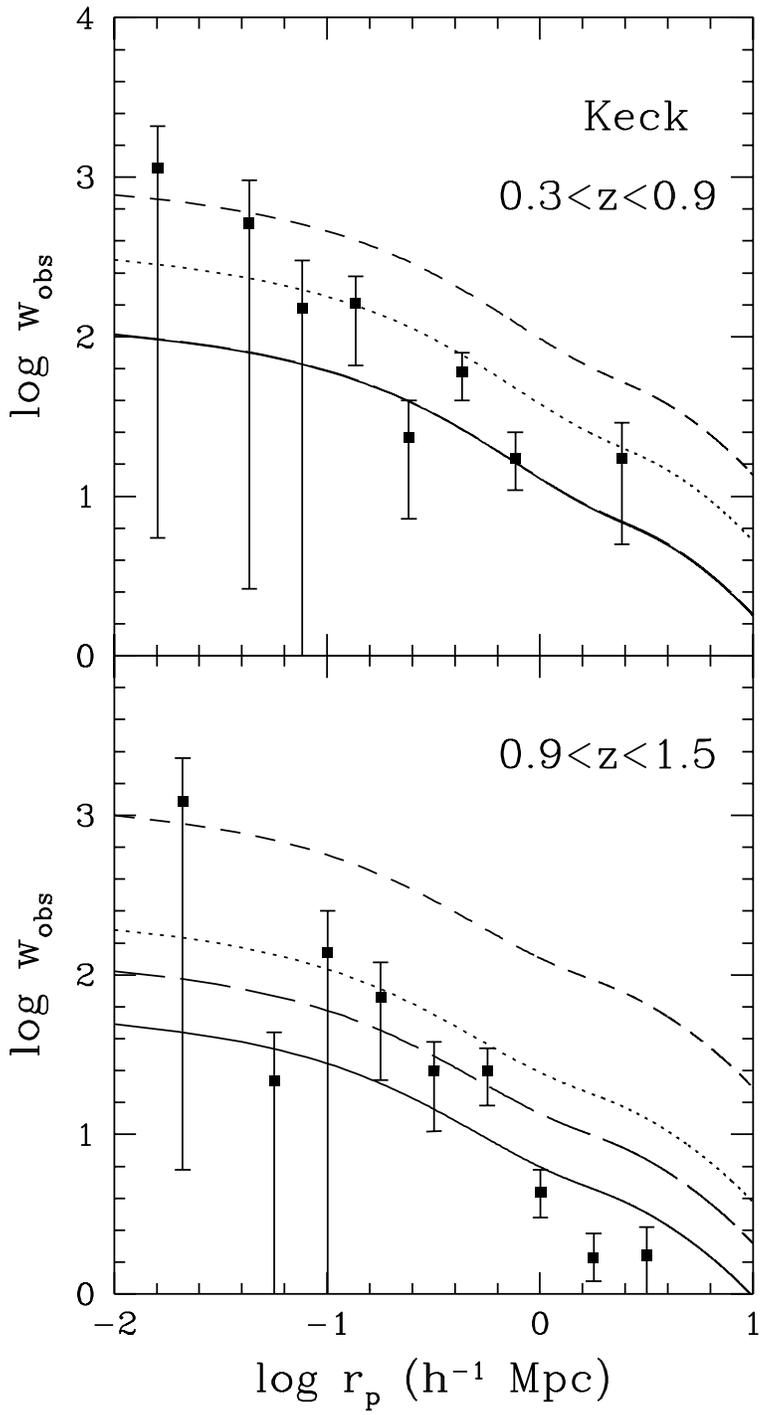
\end{document}